\documentclass[twocolumn]{aastex701}
\usepackage{graphicx} 

\newcommand\bb[1]{\mbox{\boldmath{$#1$}}}

\begin{document}

\title{Formation of Suprathermal Electron Populations in the Expanding, Turbulent Solar Wind}

\correspondingauthor{Maximilien Péters de Bonhome} 

\author[0000-0002-7493-6181,gname='Maximilien',sname='Péters de Bonhome']{Maximilien Péters de Bonhome}
\affiliation{Centre for mathematical Plasma Astrophysics, Department of Mathematics, KU Leuven, Celestijnenlaan 200B, B-3001 Leuven,
Belgium}
\affiliation{Space Physics, Solar-Terrestrial Centre of Excellence, Royal Belgian Institute for Space Aeronomy, Ringlaan 3 Avenue Circulaire,
B-1180 Brussels, Belgium}
\email[show]{maximilien.petersdebonhome@kuleuven.be}  

\author[0000-0002-7526-8154,gname='Fabio',sname='Bacchini']{Fabio Bacchini}
\affiliation{Centre for mathematical Plasma Astrophysics, Department of Mathematics, KU Leuven, Celestijnenlaan 200B, B-3001 Leuven,
Belgium}
\affiliation{Space Physics, Solar-Terrestrial Centre of Excellence, Royal Belgian Institute for Space Aeronomy, Ringlaan 3 Avenue Circulaire, B-1180 Brussels, Belgium}
\email[]{fabio.bacchini@kuleuven.be}

\author[0000-0001-5079-7941,gname='Luca',sname='Pezzini']{Luca Pezzini}
\affiliation{Centre for mathematical Plasma Astrophysics, Department of Mathematics, KU Leuven, Celestijnenlaan 200B, B-3001 Leuven,
Belgium}
\affiliation{Solar-Terrestrial Centre of Excellence--SIDC, Royal Observatory of Belgium, 1180 Brussels, Belgium}
\email[]{luca.pezzini@kuleuven.be}

\author[0000-0001-5014-7682,gname='Viviane',sname='Pierrard']{Viviane Pierrard}
\affiliation{Space Physics, Solar-Terrestrial Centre of Excellence, Royal Belgian Institute for Space Aeronomy, Ringlaan 3 Avenue Circulaire, B-1180 Brussels, Belgium}
\affiliation{Earth and Life Institute Climate Sciences ELI-C, Université Catholique de Louvain, Place Louis Pasteur 3, B-1348 Louvain-la-Neuve, Belgium}
\email[]{viviane.pierrard@aeronomie.be}

\begin{abstract}
Nonthermal features are ubiquitously observed in electron velocity distribution functions in the solar wind, yet their origin in the collisionless, turbulent, expanding solar-wind plasma remains unclear. We investigate how solar-wind expansion and Alfv\'{e}nic turbulence jointly generate and regulate these features using the first fully kinetic particle-in-cell simulation of an expanding turbulent plasma under heliospheric conditions. In our setup, expansion-driven weakening of the magnetic field adiabatically cools the plasma perpendicularly to the mean field while leaving the parallel temperature largely unchanged, driving the system toward the firehose-instability threshold. Concurrently, strongly anisotropic turbulence leads to perpendicular heating and the development of nonthermal features. Subsequently, we find that suprathermal electron populations preferentially develop in the parallel direction, forming pronounced power-law tails even under weakly compressive, highly Alfv\'{e}nic conditions, and persist despite anisotropy regulation by the firehose instability. The preferentially parallel energization suggests the involvement of parallel electric fields or resonant wave--particle interactions, rather than simple velocity-space redistribution. These results provide the first direct evidence of the emergence of nonthermal-electron features in a unified kinetic framework linking expansion, turbulence, and instabilities in the solar wind.
\end{abstract}


\section{Introduction} 

The solar wind is a weakly collisional plasma that undergoes continuous expansion as it propagates in a turbulent flow away from the Sun \citep{marsch_kinetic_2006}. The weak collisionality allows the solar wind to develop ubiquitous nonthermal features in the electron velocity distribution functions (VDFs), which usually display the presence of three components: the core, the strahl, and the halo. The core population is the bulk of the electron population and typically exhibits a Maxwellian distribution. The strahl is a field-aligned beam of suprathermal electrons that is very clearly observed in the fast solar wind. The halo is an isotropic population of suprathermal electrons that is more pronounced in the slow solar wind \citep{lazar_characteristics_2020, salem_precision_2023}. The distinction between these three populations is particularly evident in highly Alfv\'{e}nic solar wind, where the strahl and halo components can be clearly distinguished from the core \citep[e.g.,][]{bercic_coronal_2020, halekas_electrons_2020}. Strahl and halo also often exhibit suprathermal tails characterized by a power-law decay at high speeds; these tails become more pronounced at larger heliocentric distances \citep{abraham_radial_2022,salem_precision_2023,zheng_radial_2024}. Such power-law distributions are not unique to the solar wind, but are rather a ubiquitous feature of weakly collisional space and astrophysical plasmas, where they are commonly interpreted as signatures of ongoing nonthermal energization processes \citep[e.g.,][]{pierrard_kappa_2010, livadiotis_kappa_2017}. 

The radial evolution of these suprathermal components is known to regulate the heat flux of the solar wind through kinetic instabilities induced by the expansion that naturally occurs as the plasma flows away from the Sun \citep{verscharen_self-induced_2019,innocenti_collisionless_2020,micera_role_2021,coburn_regulation_2024}. In particular, this expansion induces a decrease in plasma density and magnetic-field strength, which drives the plasma toward the firehose-instability threshold by reducing the perpendicular temperature (with respect to the magnetic-field direction) while keeping the parallel temperature relatively constant, as expected from double-adiabatic predictions \citep{chew_boltzmann_1956}. The firehose instability then acts as a diffusive mechanism for suprathermal electrons that redistributes particles from higher to lower velocity-space densities (i.e., primarily from the parallel to the perpendicular direction in this context; \citealt{lopez_particle--cell_2019,verscharen_electron-driven_2022}). This, in turn, naturally regulates the heat flux of the solar wind.

The electron heat flux, being primarily carried by suprathermal electrons streaming along the magnetic field, also plays a central role in setting the large-scale ambipolar electric field that maintains quasineutrality and zero net current during the expansion. As electrons tend to escape more efficiently than ions due to their lower mass, an electric field develops along the magnetic field to maintain a balance between their relative motion \citep{lemaire_kinetic_1971}. The strength of this balancing electric field is directly influenced by the high-energy content of the electron distribution: the more pronounced the high-energy tail of the strahl population, the more efficiently electrons escape, thereby enhancing the ambipolar electric field. Exospheric models illustrate this dependence, showing that an increased suprathermal electron population leads to a stronger electrostatic potential and a more efficient acceleration of the plasma \citep{maksimovic_kinetic_1997,lamy_kinetic_2003,zouganelis_transonic_2004,peters_de_bonhome_kinetic_2025}. Through this mechanism, the ambipolar electric field contributes directly to ion acceleration and is able to account for the slow-solar-wind acceleration \citep{halekas_radial_2022, halekas_quantifying_2023}. For these reasons, understanding how electron suprathermal tails, as important contributors to the electron heat flux, are generated and evolve during the expansion is essential to determine how the solar wind is accelerated.

Turbulence is an important candidate for the generation of suprathermal tails in the solar wind. In weak-turbulence theory, small-amplitude, stochastic wave--particle interactions lead to velocity-space diffusion that can produce power-law distributions as steady-state solutions \citep{yoon_electron_2014,livadiotis_kappa_2017}. However, in strong turbulence, such as the highly Alv\'{e}nic solar wind, the nature of wave--particle interactions is more complex and may involve coherent structures, intermittency, and nonlinear effects that can enhance particle energization beyond the predictions of quasilinear theory \citep[e.g.,][]{comisso_particle_2018}. In particular, intermittent structures in strong turbulence are known to be sites of enhanced particle energization and can generate significant deviations from Maxwellian distributions \citep{bruno_intermittency_2019}. While such processes can produce high-energy populations, their direct role in forming extended power-law tails remains an open question. This issue is particularly relevant in the highly Alfv\'{e}nic and weakly compressible solar wind, where turbulent fluctuations are predominantly perpendicular to the mean magnetic field. In such a regime, particle energization is expected to be primarily perpendicular, raising the question of how suprathermal tails develop and persist in the parallel direction, where the strahl population is observed.

Previous studies have investigated the ion firehose instability in ``expanding-box'' simulations (which naturally account for solar-wind expansion) using hybrid kinetic models (i.e., with massless, fluid electrons) \citep{matteini_parallel_2006, hellinger_oblique_2008, hellinger_proton_2017}, including in the context of Alfv\'{e}nic turbulence \citep{hellinger_turbulence_2019} and critically balanced energy cascades \citep{bott_adaptive_2021}. Other studies have focused on electron-scale systems and examined the role of expansion-driven instabilities in regulating the electron heat flux, using two-dimensional simulations in homogeneous plasma conditions \citep[][]{innocenti_collisionless_2020,micera_role_2021}. These works have provided valuable insights into the behavior of these instabilities and their impact on the evolution of the electron velocity distribution. However, the interaction between turbulence, expansion, and firehose instabilities of both ions and electrons remains essentially unexplored.

In this Letter, we present the first fully kinetic simulation of an expanding, turbulent plasma in three dimensions, including the physics of ions and electrons over large (from electron to ion) spatial scales. In particular, we examine the formation of electron suprathermal tails in weakly compressive, highly Alfv\'{e}nic conditions and characterize their development in the parallel direction, providing insight into the physical processes that may govern their formation, with implications for the solar wind and other astrophysical plasmas.

\section{Numerical method and setup} 

We perform a fully kinetic particle-in-cell (PIC) simulation using \textsc{Zeltron} \citep{cerutti_simulations_2013,bacchini_fully_2022}, a relativistic PIC code for which the kinetic expanding box was implemented \citep{bacchini_particle--cell_2026}. \textsc{Zeltron} solves the fully relativistic equations of motion, which naturally recover the nonrelativistic limit when particle velocities satisfy $v \ll c$, as is the case throughout our simulation. The simulation employs coordinates in the ``expanding frame'', where expansion is not apparent and boundary conditions can be kept as periodic in all directions. The run is initiated with a guide magnetic field $\bb{B}_0$ in the $x$-direction such that $\bb{B}_0 = B_{\mathrm{g}0} \hat{\bb{x}}$. The box expands perpendicularly to the direction of the guide field (i.e., in the $y$- and $z$-directions). We define the perpendicular expansion factor as $a_\perp(t) \equiv L_\perp(t)/L_{\perp0}$ such that $a_\perp(t) =  1$ for $t < t_{\mathrm{exp}}$ and $a_\perp(t) = (1 + (t-t_{\mathrm{exp}})/\tau_{\mathrm{exp}})$ for $t \geq t_{\mathrm{exp}}$, where $\tau_{\mathrm{exp}}$ is the expansion timescale, $t_{\mathrm{exp}}$ is the onset time of the expansion, and $L_{\perp0}$ is the initial perpendicular length of the box. The parallel length of the box, $L_\parallel$, remains fixed during the expansion (i.e., $a_\parallel = 1$). Assuming a constant solar-wind velocity along $x$ and a purely radial magnetic field, both the background magnetic ($B_g$) field and the density ($n$) exhibit identical time dependence, scaling as $B_g(t) = B_{\mathrm{g}0} a_\perp^{-2}(t)$ and $n(t) = n_0 a_\perp^{-2}(t)$, where $n_0$ denotes the initial number density. More details of the expanding-box implementation are discussed in \citet{bacchini_particle--cell_2026}.

To ensure the development of turbulence during our simulation, we employ the solenoidal forcing (i.e., driven perpendicularly with respect to the background guide field) used in \citet{bott_adaptive_2021} and described in \citet{arzamasskiy_hybrid-kinetic_2019}. The forcing starts at the beginning of the run ($t=0$) and is maintained throughout the expansion. Here, however, we adopt a charge-independent forcing by applying equal force to ions and electrons (i.e., with identical strength and direction). The forcing is time-correlated via an Ornstein--Uhlenbeck process, ensuring a finite correlation time, and mimics random inertial forcing produced by an anisotropic turbulent cascade at scales larger than the simulation domain (see \citealt{arzamasskiy_hybrid-kinetic_2019} for more details).

The simulation domain size at $t=0$ is $L_x \times L_y \times L_z = 52d_i \times (13d_i)^2$ containing $768 \times 192^2$ cells, where $d_i = c/\omega_{\mathrm{p}i}$ is the ion inertial length with $\omega_{\mathrm{p}i} = (4\pi n_0 e^2/m_i)^{1/2}$ the ion plasma frequency. We choose an ion-to-electron mass ratio of $m_i/m_e = 25$ (see Section~\ref{sec:Disc} for a discussion of this choice), yielding a spatial resolution of $2.95$ cells per electron inertial length, $d_e = c/\omega_{\mathrm{p}e}$. At $t=0$, the plasma consists of $N_\mathrm{PPC} = 256$ particles per cell per species (ions and electrons) uniformly distributed across the simulation domain. For each species, the particles are drawn from a Maxwellian distribution with number density $n_0 = n_{i0} = n_{e0}$ and temperature $T_0 = T_{i0} = T_{e0}$ so that $\theta_{i0} = (m_e/m_i)\theta_{e0} = k_\mathrm{B}T_0/(m_i c^2) = 1/5000$. The magnitude of the initial background magnetic field ($B_{\mathrm{g}0}$) is set such that the initial ion Alfv\'{e}n speed $v_{\mathrm{A}i0} = B_{\mathrm{g}0}/(4\pi m_i n_0)^{1/2} = 0.02c$ (see Section~\ref{sec:Disc} for implications), and the initial ion plasma beta $\beta_{i0} = 8\pi n_0 k_\mathrm{b} T_{i0}/B_{\mathrm{g}0}^2 = 1$, consistent with typical solar wind conditions near Earth \citep{matteini_evolution_2007}. This results in an ion thermal speed $v_{\mathrm{th}i0} = \sqrt{3k_\mathrm{B}T_{i0}/m_i} \approx 0.026c $. 

\begin{figure}[t]
    \includegraphics[width=\columnwidth]{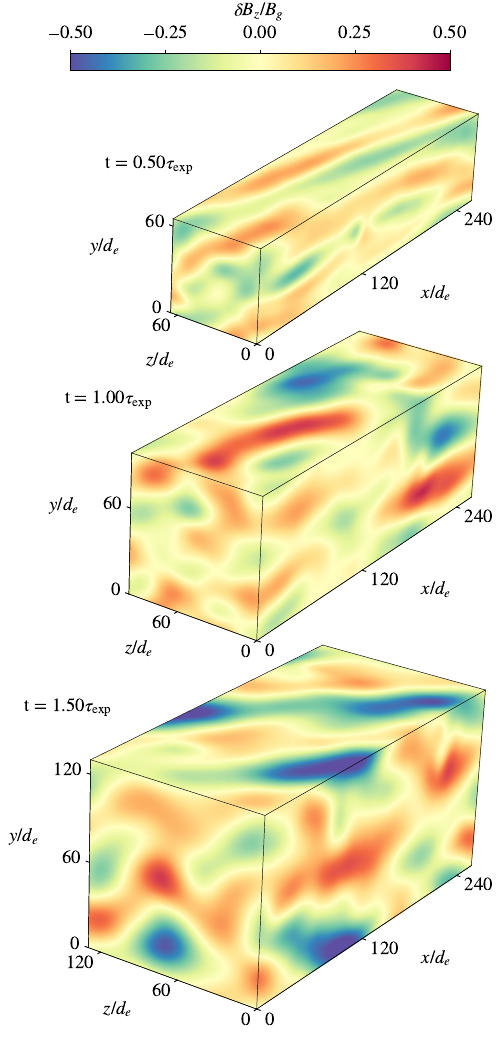}
    \caption{Volume rendering of the $z$-component of the magnetic-field fluctuations, $\delta B_z /B_g$, shown at the initial expansion time ($t = 0.5\tau_\mathrm{exp}$), mid-expansion time ($t = 1.0\tau_\mathrm{exp}$), and final expansion time ($t = 1.5\tau_\mathrm{exp}$).}
    \label{fig:EB}
\end{figure}   

\begin{figure*}[t]
    \includegraphics[width=\textwidth]{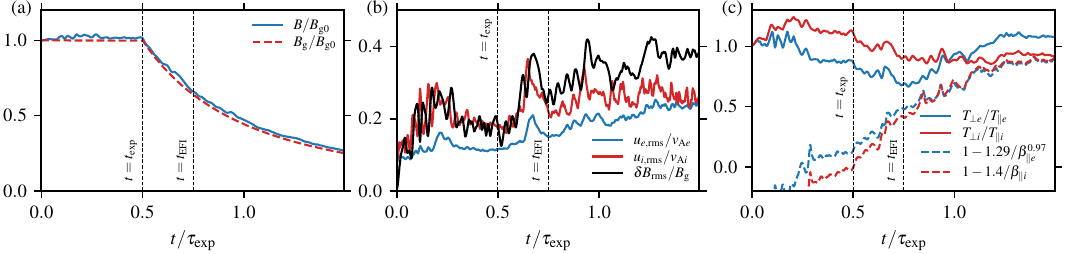}
    \caption{(a) Evolution of the box-averaged magnetic-field strength normalized by the initial background magnetic field. (b) Evolution of the root-mean-square ($\mathrm{rms}$) speed of electrons (in blue), ions (in red), and the magnetic-field fluctuations normalized by the background magnetic field in black. (c) Evolution of the average temperature ratio, $T_\perp/T_\parallel$, for electrons (in blue) and ions (in red), with the associated firehose instability thresholds plotted as corresponding dashed lines. The vertical dashed lines indicate the onset of expansion at $t_\mathrm{exp} = 0.5\tau_\mathrm{exp}$ and triggering of the electron firehose instability at $t_\mathrm{EFI} = 0.75\tau_\mathrm{exp}$.}
    \label{fig:RMA}
\end{figure*} 

Alfv\'{e}nic turbulence in our simulation is obtained through the aforementioned charge-independent forcing applied on particles. The strength of the forcing is chosen so that the ion root-mean-square velocity is $u_{i,\mathrm{rms}}/v_{\mathrm{A}i0} \approx L_\perp/L_\parallel = L_{(y,z)}/L_x$, effectively maintaining critical balance for the large-scale fluctuations constrained by the box size \citep{bott_adaptive_2021}. The simulation is first evolved without expansion until a quasi-steady state of turbulence is reached at $t_\mathrm{exp} = 5\tau_{A0}$, where $\tau_{A0} = L_\parallel/v_{\mathrm{A}i0} = 2600 \omega_{\mathrm{p}i}^{-1}$ is the initial Alfv\'{e}n crossing time. After $t_\mathrm{exp}$, the box starts expanding in the perpendicular directions at a constant rate of $q_\mathrm{exp} = 1/\tau_\mathrm{exp}$, with $\tau_\mathrm{exp} = 10\tau_{A0} = 26000 \omega_{\mathrm{p}i}^{-1}$ (so that $t_\mathrm{exp} = 0.5\tau_\mathrm{exp}$). Since the turbulent heating time, defined as the ratio of the thermal energy to the energy dissipation rate, $\tau_\mathrm{heat} \sim (3/2)T_i L_\perp / (m_i u_{i,\mathrm{rms}}^3)$, satisfies $\tau_\mathrm{heat} \gtrsim 3.75 \tau_\mathrm{exp}$ in our simulation, the thermodynamic evolution of the plasma is dominated by the expansion and the associated firehose instability.

\section{Results} 

\subsection{Global Evolution and Onset of Firehose instabilities}

The expansion of the box is displayed with the $z$-component of the turbulent magnetic field in Figure~\ref{fig:EB} from the initial expanding time $t = 0.5\tau_\mathrm{exp}$ to $t = 1.5\tau_\mathrm{exp}$.

The global evolution of various integrated quantities characterizing the system is presented in Figure~\ref{fig:RMA}, with the expansion starting at $t = t_\mathrm{exp}= 0.5\tau_\mathrm{exp}$. Figure~\ref{fig:RMA}(a) displays the normalized box-averaged magnetic-field strength (blue line), which evolves in excellent agreement with the the analytically expected scaling $B_g(t) / B_{\mathrm{g}0} = a_\perp^{-2}(t)$ (red line). A small deviation from this scaling is observed, with the magnetic-field strength remaining slightly larger than $a_\perp^{-2}(t)$. This difference arises from the contribution of magnetic-field fluctuations associated with the forced turbulence, which slows down the effective decrease of the box-averaged field as the amplitude of the fluctuations increases during the expansion phase as seen in Figure~\ref{fig:EB} and \ref{fig:RMA}(a) with the associated increase in root-mean-square magnetic-field fluctuations \cite[as observed in][]{bott_adaptive_2021}.

Figure~\ref{fig:RMA}(b) shows the evolution of the root-mean-square speeds of ions and electrons, normalized by their respective time-evolving Alfv\'{e}n speed, together with the average strength of the magnetic fluctuations normalized to the background field. Here, we define the time-evolving Alfv\'{e}n speed as $v_{\mathrm{A}i}(t) = B_g(t)/(4 \pi m_i n(t))^{1/2} = (m_i/m_e)^{1/2} v_{\mathrm{A}e}(t)$. Before the onset of expansion, the magnetic-field fluctuations satisfy $\delta B_\mathrm{rms}/B_g \sim u_{i,\mathrm{rms}}/v_{\mathrm{A}i}$, which is characteristic of Alfv\'{e}nic turbulence. During the expansion phase, the magnetic-field fluctuations increase in strength as expected with $\delta B_\mathrm{rms}/B_g \propto a_\perp(t)$ together with $u_{e,\mathrm{rms}}/v_{\mathrm{A}e} \propto a_\perp(t)$, while the ion root-mean-square velocity increases slightly slower than $a_\perp(t)$.

The evolution of the average $T_\perp/T_\parallel$ for ions and electrons is shown in Figure~\ref{fig:RMA}(c). During the nonexpanding phase, the electron population (blue line) exhibits a lower $T_\perp/T_\parallel$ than the ions (red line), with both remaining relatively close to unity. Once expansion begins, both species experience a decrease in $T_\perp/T_\parallel$, broadly following the double-adiabatic predictions (not shown), and eventually reaching close to their respective firehose instability thresholds (dashed lines). The electron firehose instability (EFI) is triggered at around $t \approx 0.75\tau_\mathrm{exp}$, as indicated by the rapid increase in $T_\perp/T_\parallel$. In contrast, the ion firehose instability lacks a clearly defined onset time, but instead develops more gradually as the expansion proceeds, with the instability threshold being reached around $t \approx 1.2\tau_\mathrm{exp}$. This difference partially reflects the initial conditions at the onset of expansion, with ions starting from a higher $T_\perp/T_\parallel$ and therefore requiring a longer time to reach the instability threshold.

\subsection{Evolution of Electron Distributions}

\begin{figure*}[t]
    \includegraphics[width=\textwidth]{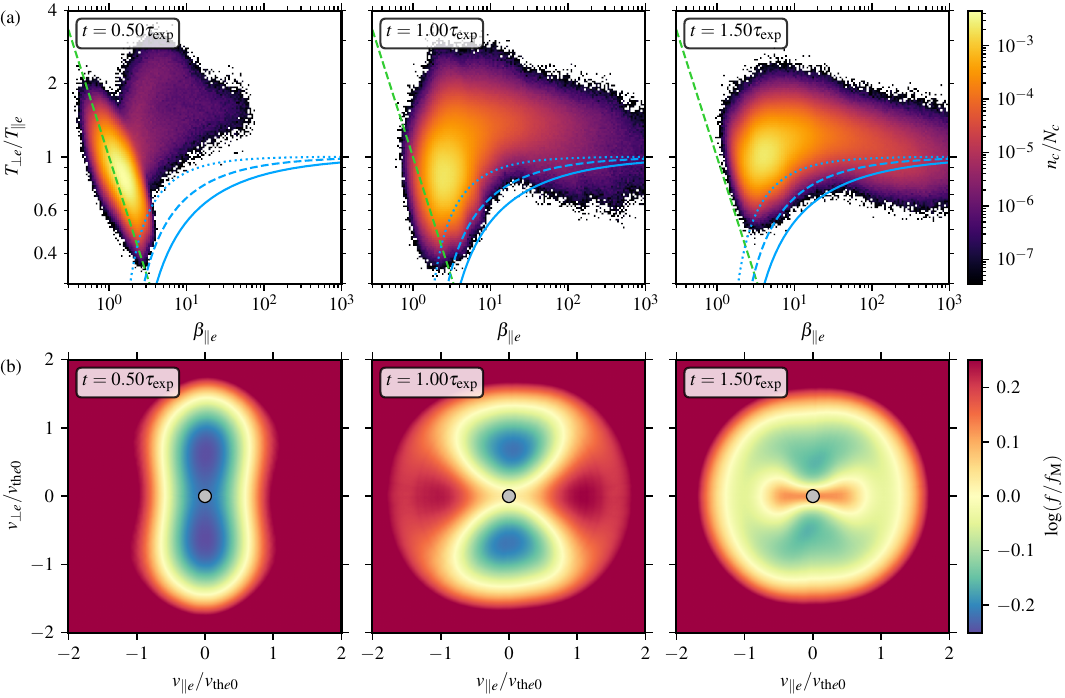}
    \caption{(a) Distribution of $(T_{\perp e}/T_{\parallel e}, \beta_{\parallel e})$ for electrons at $t = 0.5\tau_\mathrm{exp}$ (left), $t = 1.0\tau_\mathrm{exp}$ (middle), and $t = 1.5\tau_\mathrm{exp}$ (right). The distribution is scaled as the amount of cells ($n_c$) normalized by the total amount of cells ($N_c$). Isocontours of the theoretical oblique EFI are plotted as blue lines with $T_{\perp e}/T_{\parallel e} = 1 - 1.29/\beta_{\parallel e}^{0.97}$ (dotted), $T_{\perp e}/T_{\parallel e} = 1 - 1.32/\beta_{\parallel e}^{0.61}$ (dashed), and $T_{\perp e}/T_{\parallel e} = 1 - 1.36/\beta_{\parallel e}^{0.47}$ (solid) associated with growth rates of $\gamma = 0.001, 0.1$ and $0.2 \Omega_e$ respectively \citep{gary_resonant_2003} with $\Omega_e = eB/(m_e c)$ and $e$ is the electron charge. The green dashed line corresponds to the double-adiabatic expansion prediction ($T_{\perp e}/T_{\parallel e} = \beta_{\parallel e}^{-1}$). (b) Logarithmic ratio between the electron VDFs ($f$) and a reference Maxwellian VDF ($f_M$) in $(v_{\perp e}, v_{\parallel e})$-space at $t = 0.5\tau_\mathrm{exp}$ (left), $t = 1.0\tau_\mathrm{exp}$ (middle), and $t = 1.5\tau_\mathrm{exp}$ (right).}
    \label{fig:BD}
\end{figure*}  

We now examine the distribution of $(T_{\perp}/T_{\parallel}, \beta_\parallel)$ for electrons, shown in Figure~\ref{fig:BD}(a). At the onset of expansion, $t =t_\mathrm{exp}= 0.5\tau_\mathrm{exp}$, the electron population is distributed predominantly along the double-adiabatic prediction (green line). As the expansion proceeds, the distribution generally evolves along the double-adiabatic trajectory toward the firehose-instability thresholds (blue lines), with a significant fraction of the plasma becoming unstable at $t = 1.0\tau_\mathrm{exp}$. At later times, $t = 1.5\tau_\mathrm{exp}$, the main part of the distribution returns close to isotropy, consistent with a state near marginal stability.

To qualitatively assess the role of the firehose instability in shaping the electron velocity distribution functions (VDFs), Figure~\ref{fig:BD}(b) presents the evolution of the VDFs, restricted to the range $\pm 2v_{\mathrm{th}e0}$, as a logarithmic ratio to the initial reference Maxwellian. The saturated red regions of velocity space correspond to large deviations at higher velocities, where the logarithmic contrast is strongest; the color scale is therefore limited to highlight differences in the central part of the distribution. At $t = 0.5\tau_\mathrm{exp}$, the core of the electron VDF is already anisotropic due to our turbulence-driving method, with a more extended deficit of particles in the perpendicular direction than in the parallel direction relative to the reference Maxwellian (see next section for further discussion on this point). At later times, this deficit becomes confined to the perpendicular direction, while the parallel direction becomes more populated than the reference Maxwellian. This behavior, characteristic of the firehose instability, has been previously reported for ions \citep[e.g.,][]{hellinger_proton_2017,bott_adaptive_2021}, but, to our knowledge, has not been observed for electrons in a similar context. By $t = 1.5\tau_\mathrm{exp}$, the contrast between the parallel and perpendicular directions is reduced, as indicated by the smaller difference between the two directions. This is consistent with the plasma returning close to isotropy and approaching marginal stability, in agreement with previous proton studies \citep{bott_adaptive_2021}.

\begin{figure*}[t]
    \includegraphics[width=\textwidth]{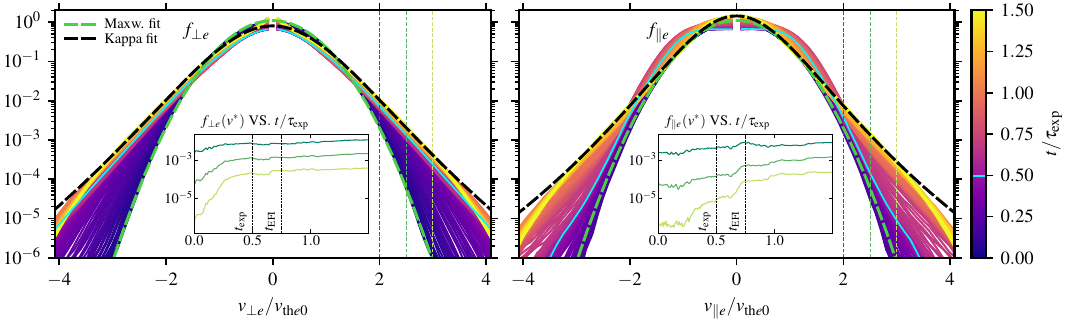}
    \caption{Time evolution of the perpendicular ($f_e(v_{\perp e})$) and parallel ($f_e(v_{\parallel e})$) electron VDFs. The dashed green line corresponds to a Maxwellian fit at $t = 0$, while the dashed black line corresponds to a Kappa distribution fit \citep[e.g.,][]{salem_precision_2023} at $t = 1.5\tau_\mathrm{exp}$ with parameter $\kappa = 6.2$ for $f_e(v_{\perp e})$ and $\kappa = 5$ for $f_e(v_{\parallel e})$. The solid blue line indicates the electron VDF at $t = 0.5\tau_\mathrm{exp}$ (onset time of expansion). The insets show the time evolution of the electron VDF at a fixed speed $v^* = \{2.0, 2.5, 3.0\} v_{\mathrm{th}e0}$ associated with the vertical dashed lines displayed in their respective main panel.}
    \label{fig:VDF_cuts}
\end{figure*}  

To better characterize the deviations from a Maxwellian, Figure~\ref{fig:VDF_cuts} presents one-dimensional cuts of the electron VDFs along the parallel and perpendicular directions, using the full distributions rather than their ratio to a reference Maxwellian. We also highlight the distribution at the onset of expansion (solid blue line), a Maxwellian (dashed green line) and a Kappa (dashed black line) fit of the distribution at the initial and final simulation time, respectively. The left panel then shows the distribution along the perpendicular direction: this displays a clear heating of the electrons already before the onset of expansion, associated with our choice of forcing, with well-developed suprathermal tails already present at $t = 0.5\tau_\mathrm{exp}$. During the subsequent evolution, no significant further qualitative change is observed in the perpendicular direction, and the tails remain relatively stable, aside from a brief enhancement around $t \approx 0.67\tau_\mathrm{exp}$.

The right panel of Figure~\ref{fig:VDF_cuts} shows the distribution along the parallel direction. At $t = 0.5\tau_\mathrm{exp}$, no significant suprathermal component is observed in the parallel direction, except for a slight departure from a Maxwellian at higher energies. As time progresses, a pronounced suprathermal tail develops, which is well-fitted by a Kappa distribution at $t = 1.5\tau_\mathrm{exp}$ with parameter $\kappa = 5$. The emergence of this tail occurs abruptly around $t \approx 0.67\tau_\mathrm{exp}$ with the plasma approaching marginal stability slightly before the onset of EFI at $t = t_{\mathrm{EFI}} = 0.75\tau_\mathrm{exp}$, as observed from the inset of the right panel of Figure~\ref{fig:VDF_cuts}.

\section{Discussion}
\label{sec:Disc}

We have conducted the first fully kinetic simulations of expanding, turbulent plasma under heliospheric conditions, including the physics of ions and electrons. Our simulations employ a solenoidal driver to maintain Alfv\'{e}nic turbulence active throughout the run, and an expanding-box algorithm to track plasma expansion for a full expansion time after the turbulence has reached a quasi-steady state. Our main result is the observation of the emergence of a clear nonthermal electron population, particularly in the parallel direction, driven and regulated by the interplay of turbulence, expansion, and firehose instability.

The fact that the emergence of high-energy electrons in the parallel direction precedes the onset of the EFI indicates that the mechanism responsible for the formation of the parallel suprathermal tail is not simply a consequence of a global state of instability, and therefore requires further investigation. In particular, while the plasma may already be locally unstable to the EFI in some regions of the domain at this stage, these localized events do not immediately translate into a sustained global response of the system. Instead, the later marginally stable state appears to operate on a distribution that has already been significantly modified.

The rapid and pronounced increase in the electron parallel energy suggests the activation of a distinct process as the plasma approaches marginal stability. The sharpness of this transition (as seen in insets of Figure \ref{fig:VDF_cuts}) points to a mechanism that becomes particularly efficient under specific conditions, rather than a gradual, background diffusive process. This behavior may be facilitated by intermittent or localized structures, which can enhance particle energization before the state of marginal stability is reflected in the perpendicular to parallel temperature ratio.

Importantly, this behavior cannot be attributed solely to velocity-space redistribution from perpendicular to parallel directions. No clear evidence of a corresponding depletion of the perpendicular suprathermal population is observed, as would be expected if pitch-angle scattering were dominant. This instead points toward a mechanism involving direct acceleration in velocity space, capable of increasing particle energy rather than merely redistributing it. Within the context of our setup, this energization appears to be preferentially efficient in the parallel direction, suggesting the involvement of processes associated with parallel electric fields or resonant wave–particle interactions. Such a process enables the buildup of a suprathermal tail and drives the system closer to the instability threshold.

The main implication is that, prior to the onset of the EFI, the plasma remains only weakly constrained and can develop both significant temperature anisotropy and pronounced non-Maxwellian features, including parallel suprathermal tails. Although localized regions may already experience EFI-driven dynamics, it is only once the instability reaches a global, marginally stable state that it effectively regulates the system. At that stage, the EFI constrains further departures from isotropy while preserving the suprathermal tails, indicating that their formation is largely decoupled from the state of marginal stability. 

In this context, it is worth noting that the bi-Maxwellian isocontours \citep{gary_resonant_2003} used as a reference for the EFI in this work are likely underestimated. \cite{shaaban_firehose_2019} showed that the dominant branch of the EFI—the oblique resonant mode—is enhanced by the presence of suprathermal electron tails, leading to a more rapid approach to marginal stability.

A key limitation of this study lies in the formation of non-Maxwellian electron VDFs (only in the perpendicular direction) prior to the onset of expansion. This feature arises from the requirement that a fully developed turbulent state be established beforehand, which leads to significant turbulent heating and the formation of perpendicular suprathermal tails. For comparison, the simulation of \cite{bott_adaptive_2021} did not exhibit turbulent heating, as hyper-resistivity was introduced to account for their fluid-electron treatment. As a result, their system remained closer to a Maxwellian state prior to expansion. In contrast, our system departs from a Maxwellian state before expansion begins, and the subsequent development of a parallel suprathermal tail may be influenced by the prior existence of perpendicular suprathermal populations. In other words, the responsible mechanism may not operate, or may operate differently, in plasmas that do not already exhibit strong perpendicular energization. However, we believe our results qualitatively hold for the following reasons. First, as we discussed, the acceleration in the parallel direction appears to be caused by direct acceleration and/or resonant absorption, without necessarily involving any perpendicular dynamics. It is therefore plausible to expect that an isotropic initial distribution would also develop nonthermal parallel features. Second, in the solar wind, the turbulent heating, expansion-driven anisotropy, and the approach to marginal stability are expected to occur simultaneously rather than sequentially. Under such conditions, the generation of perpendicular suprathermal tails and the onset of firehose-driven fluctuations would likely coevolve. It is therefore plausible that a similar mechanism responsible for generating suprathermal tails of electrons in the parallel direction would also operate under more realistic conditions. This hypothesis remains to be tested in future work, which will investigate the coupled evolution of turbulence, expansion, and kinetic instabilities starting from nearly Maxwellian initial conditions at the onset of expansion. Finally, we can also argue that while our results may depend on the plasma state at the onset of expansion, this state is not necessarily unphysical. In fact, the state in which expansion starts (with a perpendicular anisotropy) is just one of many possible states which in reality would depend on where the plasma parcel that starts expanding has originated. Therefore, this state is as valid as any other as far as initial conditions are concerned. Nevertheless, more experiments with different initial conditions are needed to draw general conclusions; but to the best of our knowledge, this is the first time that electron nonthermal features are observed arising in fully kinetic simulations of turbulent, expanding solar plasmas.

For computational feasibility, both the ion-to-electron mass ratio ($m_i/m_e = 25$) and the light-to-Alfvén speed ratio ($c/v_{\mathrm{A}i} \gtrsim 50$) used in our simulations are reduced compared to typical solar-wind conditions. This approach (dictated by computational demands and customary in PIC simulations) decreases scale separation and may affect wave--particle resonance conditions. This effect may already be reflected in our results, as evidenced by the slower increase in ion $T_\perp/T_\parallel$ during expansion compared to electrons, suggesting that ions are partially influenced by electron-scale processes. However, expansion-driven instabilities are primarily governed by dimensionless parameters, such as the parallel plasma-$\beta$ and temperature anisotropy, which are initialized with solar-wind-like values, ensuring that the system remains in a physically relevant regime. Moreover, the properties of the fast magnetosonic/whistler modes, which can be generated by the dominant EFI among other mechanisms, are only weakly affected provided that $m_i/m_e \gtrsim 10$ and $c/v_{\mathrm{A}i} \gtrsim 10$ \citep{verscharen_dependence_2020}, conditions that are well satisfied here. We therefore do not expect these numerical constraints to qualitatively affect the electron-scale dynamics or the development of suprathermal tails discussed in this work. Future work will explore a broader parameter space, including more realistic mass ratios, to further assess the robustness of these results.

\begin{acknowledgments}
We would like to thank Archie Bott, Daniel Verscharen, and Daniel Gro\v{s}elj for useful discussions throughout the development of this work.
F.B. acknowledges support from the FED-tWIN programme (profile Prf-2020-004, project ``ENERGY''), issued by BELSPO, and from the FWO Junior Research Project G020224N granted by the Research Foundation -- Flanders (FWO).
L.P. acknowledges support from the Fonds voor Wetenschappelijk Onderzoek PhD fellowship (grant no. 11PCB24N).
The computational resources and services used in this work were provided by the VSC (Flemish Supercomputer Center), funded by the Research Foundation Flanders (FWO) and the Flemish Government, department EWI.
\end{acknowledgments}

\begin{contribution}

\end{contribution}

\software{\textsc{Zeltron} (\citealt{cerutti_simulations_2013,bacchini_fully_2022,bacchini_particle--cell_2026})}

\bibliographystyle{aasjournalv7}
\bibliography{PIC_EB_supra.bib}{}

\end{document}